\documentclass[11pt]{article}

\usepackage[latin1]{inputenc}

\usepackage{amstext,amsmath,amssymb,amsfonts,bbm,amsmath}
\usepackage{extarrows}
\usepackage{xcolor} 
\usepackage{hyperref}
\hypersetup{
    colorlinks=false,
    menubordercolor=red,
    linkbordercolor=blue
}
\usepackage{authblk}
\usepackage{caption} 
\usepackage{epsfig} 
\usepackage{color} 
\usepackage{graphicx} 
\usepackage{braket} 
\usepackage{slashed} 
\usepackage{dsfont} 
\usepackage{amsthm}  
\usepackage{multirow}

\usepackage{cite}

\allowdisplaybreaks[4]

\usepackage{psfrag}
\usepackage{tikz} 
\usetikzlibrary{arrows}
\usetikzlibrary{plotmarks}
\usetikzlibrary{external}
\usetikzlibrary{patterns}
\usepackage{pgfplots}
\pgfplotsset{compat=1.9}
\usetikzlibrary{patterns}
\usetikzlibrary{calc}

\usepackage{float}  
\usepackage{subfig}
\usepackage[lmargin=50pt,rmargin=60pt,tmargin=60pt,bmargin=65pt]{geometry}

\newcommand{\be}{\begin{equation}}
\newcommand{\ee}{\end{equation}} 

\newcommand{\f}{\frac}

\let\a=\alpha     
\let\z=\zeta        
    \let\n=\nu

     \let\X=F

\newcommand{\cN}{\mathcal{N}}

\newcommand{\gt}{\tilde{g}}
\newcommand{\rt}{\tilde{r}}

\newcommand{\mba}{\mathbf{a}}
\newcommand{\mbb}{\mathbf{b}}
\newcommand{\mbc}{\mathbf{c}}
\newcommand{\mbd}{\mathbf{d}}
\newcommand{\mbe}{\mathbf{e}}
\newcommand{\mbf}{\mathbf{f}}
\newcommand{\mbg}{\mathbf{g}}
\newcommand{\mbh}{\mathbf{h}}
\newcommand{\mbm}{\mathbf{m}}
\newcommand{\mbn}{\mathbf{n}}

\allowdisplaybreaks[4]

\numberwithin{equation}{section}

\theoremstyle{remark}

\begin{document}

\title{\bf Addendum:\\ Long-range multi-scalar models at three loops}

\author[1]{Dario Benedetti}
\author[2]{Razvan Gurau}
\author[3]{Sabine Harribey \footnote{Corresponding author}}

\affil[1]{\normalsize\it 
CPHT, CNRS, \'Ecole polytechnique, Institut Polytechnique de Paris, 91120 Palaiseau, France
\authorcr \hfill }
	
\affil[2]{\normalsize\it 
Heidelberg University, Institut f\"ur Theoretische Physik, Philosophenweg 19, 69120 Heidelberg, Germany
\authorcr \hfill
}

\affil[3]{\normalsize \it 
 School of Theoretical Physics, Dublin Institute For Advanced Studies, 10 Burlington Road, Dublin, Ireland
  \authorcr \hfill
  \authorcr
  emails: dario.benedetti@polytechnique.edu, gurau@thphys.uni-heidelberg.de, sabine.harribey@su.se
 \authorcr \hfill }

\date{}

\maketitle

\hrule\bigskip

\begin{abstract}
We correct the computation of one Feynman diagram in the three-loop beta functions for the long-range quartic multi-scalar model, originally presented in (2020 J. Phys. A: Math. Theor. 53 445008) [arXiv:2007.04603]. The correction requires the use of a different method than in the original paper, and we give here full details about the method. We then report the updated numerics for critical exponents of the Ising model, vector model, cubic model and bifundamental model. 

Mathematica files for the numerical evaluation of the corrected diagram are provided in ancillary.
\end{abstract}

\hrule\bigskip

\tableofcontents

\section{Introduction}
\label{sec:introduction}

In \cite{Benedetti:2020rrq}, the beta functions for a $d$-dimensional long-range multi-scalar model with long-range exponent $\zeta=(d+\epsilon)/4$ and generic quartic interactions were computed to three-loop order, thus improving on old results obtained at two loops for specific interactions \cite{Fisher:1972zz,Yamazaki:1977pt}.
Unfortunately, a mistake was made in the computation of one integral, there denoted by $I_4$, corresponding to the tetrahedron Feynman diagram. Unlike the other integrals, $I_4$ was computed by Gegenbauer polynomials technique, but a formula valid only for $\zeta=(d-2)/2$ was incorrectly used for generic $\zeta$ (equation (C.46) in \cite{Benedetti:2020rrq}), thus leading to a wrong result.
The mistake could in principle be cured by expressing the Gegenbauer polynomials $C_n^\zeta(x)$ as a linear combination of $C_n^{(d-2)/2}(x)$ (e.g. \cite{Kotikov:2000yd}).
However, such method leads to multiple series that are difficult to evaluate.

Here, we will instead correct the mistake by evaluating $I_4$ via a Mellin-Barnes representation, that is the method used in \cite{Benedetti:2020rrq} for all the other three-loops integrals.
For $I_4$ the resulting series are much more complicated than for the other integrals, and we ultimately have to resort to a numerical evaluation. As a cross-check of our result, we perform the same computation also by direct numerical integration, after decomposition into Hepp sectors.

The three-loop beta functions find applications in the computation of fixed points and critical exponents for several models with various symmetry restrictions on the quartic interaction. Several of these applications were reported in \cite{Benedetti:2020rrq},\footnote{Other applications have appeared more recently, in \cite{Behan:2023ile,Rong:2024vxo,Li:2024uac}.} and we revisit them in Section~\ref{sec:appl}, correcting the numerical results in light of the corrected integral $I_4$.

\bigskip
\bigskip

\section{Three-loop beta functions of the long-range multi-scalar model}
\label{sec:model}
The long-range multi-scalar model with quartic interactions in dimension $d$ is defined by the action:
\begin{align} \label{eq:action}
		S[\phi]  \, &= \,  \int d^dx \, \bigg[ \frac{1}{2} \phi_\mba(x) ( - \partial^2)^{\zeta}\phi_{\mba}(x) +\frac{1}{2}\, \kappa_{\mba \mbb}\phi_{\mba}(x) \phi_{\mbb}(x) + 
		\frac{1}{4!} \, \lambda_{\mba \mbb \mbc \mbd}
		\phi_{\mba}(x) \phi_{\mbb}(x) \phi_{\mbc}(x) \phi_{\mbd}(x) \bigg] \, ,
\end{align}
where the indices take values from 1 to $\cN$, and a summation over repeated indices is implicit.
The  coupling $\lambda_{\mba\mbb\mbc\mbd}$ and the mass parameter $\kappa_{\mba\mbb}$ are symmetric tensors, thus corresponding in general to $\binom{\cN+3}{4}$ and $\tfrac{\cN(\cN+1)}{2}$ couplings, respectively.
The model is ``long range'' for $0< \zeta < 1$, due to the non-integer power of the Laplacian, which can be defined as the Fourier transform of $p^{2\zeta}$.  
From now on the dimension $d$ is fixed to be smaller than (and not necessarily close to) four. 

The beta functions are given by:
\begin{align}
\beta^{(4)}_{\mba \mbb \mbc \mbd} &= -\epsilon \gt_{\mba \mbb \mbc \mbd} + \alpha_{D}\left(\gt_{\mba \mbb \mbe \mbf}\gt_{\mbe \mbf \mbc \mbd} + 2 \textrm{ terms} \right)  + \alpha_{S}\left(\gt_{\mba \mbb \mbe \mbf}\gt_{\mbe \mbg \mbh \mbc}\gt_{\mbf \mbg \mbh \mbd}+ 5 \textrm{ terms}\right) \crcr 
& \quad   + \, \alpha_{U} (\gt_{\mba \mbe \mbf \mbg} \gt_{\mbb \mbe \mbf \mbh} \gt_{\mbg \mbm \mbn \mbc} \gt_{\mbh \mbm \mbn \mbd} + 5 \textrm{ terms} )  + \, \alpha_{T} (\gt_{\mba \mbb \mbe \mbf} \gt_{\mbe \mbg \mbh \mbm} \gt_{\mbf \mbg \mbh \mbn} \gt_{\mbm \mbn \mbc \mbd} + 2 \textrm{ terms} )   \crcr
& \quad    + \, \alpha_{I_1} (\gt_{\mba \mbb \mbe \mbf} \gt_{\mbe \mbg \mbh \mbm} \gt_{\mbf \mbg \mbn \mbc} \gt_{\mbh \mbm \mbn \mbd} + 11 \textrm{ terms} ) + \,\alpha_{I_2}(\gt_{\mba \mbb \mbe \mbf} \gt_{\mbe \mbg \mbh \mbc} \gt_{\mbf \mbm \mbn \mbd} \gt_{\mbg \mbh \mbm \mbn} + 5 \textrm{ terms} )  \crcr
& \quad  + \, \alpha_{I_3} (\gt_{\mba \mbb \mbe \mbf}\gt_{\mbh \mbm \mbn \mbf}\gt_{\mbh \mbm \mbn \mbg}\gt_{\mbg \mbe \mbc \mbd} +2 \textrm{ terms} ) + \, \alpha_{I_4} ( \gt_{\mba \mbe \mbm \mbh}\gt_{\mbb \mbe \mbf \mbn}\gt_{\mbc \mbf \mbm \mbg}\gt_{\mbd \mbg \mbn \mbh} ) \, . 
\label{eq:beta_abcd_alpha}
\end{align}

The coefficients alpha were computed in \cite{Benedetti:2020rrq}, and we report them here, with the exception of $\alpha_{I_4}$, which we will correct in the next section:
\begin{align}
& \alpha_{D} \, = \, 1 +\frac{\epsilon}{2}\big[\psi(1)-\psi(\tfrac{d}{2}) \big]+\frac{\epsilon^2}{8}\left[\left(\psi(1)-\psi(\tfrac{d}{2})\right)^2+ \psi_1(1)-\psi_1(\tfrac{d}{2})\right] \,, \crcr
&\alpha_{S} \, = \,  2\psi( \tfrac{d}{4} ) - \psi( \tfrac{d}{2})-\psi(1)   +\frac{\epsilon}{4}\Big[\left[2\psi(\tfrac{d}{4})-\psi(\tfrac{d}{2})-\psi(1)\right]
\left[3\psi(1)-5\psi(\tfrac{d}{2})+2\psi(\tfrac{d}{4})\right]   \crcr
& \qquad \; + 3\psi_1(1) + 4\psi_1(\tfrac{d}{4})-7\psi_1(\tfrac{d}{2})  -4 J_0(\tfrac{d}{4}) \Big] \, , \crcr
& \alpha_{U} = \alpha_{I_2} =\,  - \psi_1(1)-\psi_1(\tfrac{d}{4})+2\psi_1( \tfrac{d}{2})
+ J_0(\tfrac{d}{4})\, , \crcr
& \alpha_{T} \, = \, \frac{1}{2}\Big[2\psi(\tfrac{d}{4}) - \psi(\tfrac{d}{2})-\psi(1) \Big]^2 + \frac{1}{2}  \psi_1(1)+ \psi_1(\tfrac{d}{4}) - \frac{3}{2} \psi_1(\tfrac{d}{2}) - \, J_0(\tfrac{d}{4}) \, , \crcr
& \alpha_{I_1} \, = \, \frac{3}{2}\left[2\psi( \tfrac{d}{4} ) - \psi( \tfrac{d}{2})-\psi(1)\right]^2
 + \frac{1}{2} \psi_1(1) 
-\frac{1}{2}\psi_1(\tfrac{d}{2})
\crcr
& \alpha_{I_3} \, = \,\frac{\Gamma(-\tfrac{d}{4})\Gamma(\tfrac{d}{2})^2}{3 \, \Gamma( \tfrac{3d}{4})}\, , 
 \label{eq:alphas}
\end{align}
with $\psi_i$ being the polygamma function of order $i$ and $J_0$ the sum
\begin{equation}
J_0(\tfrac{d}{4})=\frac{1}{\Gamma(\tfrac{d}{4})^2}\sum_{n \geq 1}\frac{\Gamma(n+\tfrac{d}{2})\Gamma(n+ \tfrac{d}{4})^2}{n(n!)\Gamma(\tfrac{d}{2}+2n)}\Big[2\psi(n+1)-\psi(n)-2\psi(n+\tfrac{d}{4})-\psi(n+\tfrac{d}{2})+2\psi(\tfrac{d}{2}+2n)\Big] \,.
\end{equation}
%

\section{Computation of $I_4$}

\begin{figure}[htbp]
    \centering
    \includegraphics[scale=1]{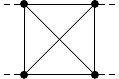}
    \caption{Three-loop Feynman graph $I_4$.}
    \label{fig:I4}
\end{figure}

In this section we correct the computation of the coefficient $\alpha_{I_4}$ given by:
\begin{equation}
    \alpha_{I_4} \, =\, \epsilon  (4\pi)^{3d/2}\Gamma(\tfrac{d}{2})^3 \,  3I_4 \, ,
\end{equation}
where $I_4$ is the dimensionless amplitude of the three-loop Feynamn graph represented in figure \ref{fig:I4}.

As this diagram does not contain divergent subgraphs, it diverges as a simple pole in $\epsilon$ and we are only interested in determining the residue at the pole. The ultraviolet divergence arises when all three-loop momenta are large, and its coefficient is independent of the chosen IR regularization.

\subsection{Computation with Mellin-Barnes representation of the kite integral}

We will compute the $I_4$ integral setting two external momenta to zero, thus having only one momentum flowing through the graph.
We can then write:
\begin{equation}
I_4 =\mu^{3\epsilon} \int \, \frac{d^d q_1}{(2\pi)^d} \frac{d^d q_2}{(2\pi)^d}  \frac{d^d q_3}{(2\pi)^d}  \frac{1}{q_1^{2\zeta}q_2^{2\zeta}(q_3+p)^{2\zeta}(q_1+q_2)^{2\zeta}(q_1+q_3)^{2\zeta}(q_1+q_2+q_3)^{2\zeta}}  \;.
\end{equation}

\begin{figure}[htb]
    \centering
    \includegraphics[width=0.5\linewidth]{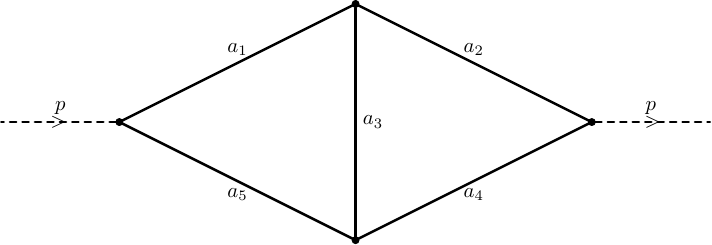}
    \caption{Generic kite graph with external momentum $p$ and powers $a_i$ of the propagators.}
    \label{fig:kite}
\end{figure}

Let us first consider only the integral over $q_1$ and $q_2$:
\begin{equation}
    \int \, \frac{d^d q_1}{(2\pi)^d} \frac{d^d q_2}{(2\pi)^d}   \frac{1}{q_1^{2\zeta}q_2^{2\zeta}(q_1+q_2)^{2\zeta}(q_1+q_3)^{2\zeta}(q_1+q_2+q_3)^{2\zeta}} \, .
\end{equation}
This is the integral of a kite diagram (see figure \ref{fig:kite}), with external momentum $q_3$ and all propagators to power $\zeta$. By power counting we notice that this integral is finite. We can thus write:
\begin{equation}
     \int \, \frac{d^d q_1}{(2\pi)^d} \frac{d^d q_2}{(2\pi)^d}   \frac{1}{q_1^{2\zeta}q_2^{2\zeta}(q_1+q_2)^{2\zeta}(q_1+q_3)^{2\zeta}(q_1+q_2+q_3)^{2\zeta}}=\frac{K_0}{q_3^{2(5\zeta-d)}} \, ,
\end{equation}
with a finite coefficient $K_0$ to be determined.
We then have:
\begin{equation}
    I_4=\mu^{3\epsilon} K_0 \int  \frac{d^d q_3}{(2\pi)^d} \frac{1}{q_3^{2(5\zeta-d)}(q_3+p)^{2\zeta}}\,.
\end{equation}

The remaining integral over $q_3$ is a standard one-loop type integral and it can be performed by means of the formula:
\begin{equation}
\int \frac{d^d k}{(2\pi)^d}\frac{1}{k^{2\alpha}(k+p)^{2\beta}}=\frac{1}{(4\pi)^{d/2}}\frac{\Gamma(d/2-\alpha)\Gamma(d/2-\beta)\Gamma(\alpha+\beta-d/2)}{\Gamma(\alpha)\Gamma(\beta)\Gamma(d-\alpha-\beta)}\frac{1}{|p|^{2\alpha+2\beta-d}} \, .
\label{eq:1loopint}
\end{equation}

We then obtain:
\begin{align}
    I_4&=K_0\frac{\Gamma(d/2-\zeta)\Gamma(3d/2-5\zeta)\Gamma(6\zeta-3d/2)}{(4\pi)^{d/2}\Gamma(\zeta)\Gamma(5\zeta-d)\Gamma(2d-6\zeta)} \crcr
    &=\frac{2K_0}{3\epsilon (4\pi)^{d/2}\Gamma(d/2)}  +\mathcal{O}(1)\, .
\end{align}

We now have to compute $K_0$, which corresponds to a kite integral. We will first compute it in the general case with arbitrary powers of the propagators and then apply it to our case.


\subsubsection{Generic Kite formula}

Many attempts have been made in the literature to compute the kite integral, as shown on figure \ref{fig:kite}, see for example \cite{Gorishnii:1984te,Grozin:2012xi,Kotikov:2018wxe}. Most of these attempts however considered either massive propagators or integer powers of the propagators \cite{Adams:2016xah,Moch:2001zr}, which does not apply to our case. Generic non-integer powers of the propagator will lead to more complicated functions: for example, \cite{Mikhailov:2018udp} gives the result as double hypergeometric series. Here we will use a representation with two Mellin-Barnes parameters \cite{Davydychev:1995mq}: 

\begin{align}
I(a_1,a_2,a_3,a_4,a_5)&=\frac{p^{2(d-\sum_i a_i)}}{(4\pi)^d \Gamma(d-a_2-a_3-a_4)\Gamma(a_2)\Gamma(a_3)\Gamma(a_4)}\int [ds][dt] \Gamma(-s)\Gamma(-t) \crcr
& \qquad \frac{\Gamma(\tfrac{d}{2}-a_2-a_3-s)\Gamma(\tfrac{d}{2}-a_5+s)}{\Gamma(a_5-s)}\frac{\Gamma(\tfrac{d}{2}-a_4-a_3-t)\Gamma(\tfrac{d}{2}-a_1+t)}{\Gamma(a_1-t)}\crcr
& \qquad \frac{\Gamma(a_3+s+t)\Gamma(a_1+a_5-\tfrac{d}{2}-s-t)\Gamma(a_2+a_3+a_4-\tfrac{d}{2}+s+t)}{\Gamma(d-a_1-a_5+s+t)} \, ,
\label{eq:kite}
\end{align}
where the contours are chosen depending on the values of the $a_i$ so that all Gamma functions have positive argument.  For the applications we are interested in, this will imply that ${\rm Re}(s)$ and ${\rm Re}(t)$ lie inside the interval $(-1,0)$, and eventually satisfy other constraints such as ${\rm Re}(s+t)>-d/4$.
By closing the integration contour of $t$ and $s$ to the right, and using the residue theorem, we can express this integral in terms of double-sums. 

We first close the contour for $t$. There are three families of poles located at $t=n_1$, $t=d/2-a_4-a_3+n_1$ and $t=a_1+a_5-d/2-s+n_1$ for $n_1$ a non-negative integer.

We obtain 
\be
I(a_1,a_2,a_3,a_4,a_5)=\frac{p^{2(d-\sum_i a_i)}}{(4\pi)^d \Gamma(d-a_{234})\Gamma(a_2)\Gamma(a_3)\Gamma(a_4)}\left(I_a+I_b+I_c\right) \;,
\ee
with:
\begin{align}
    I_a&= \sum_{n_1\geq 0}\frac{(-1)^{n_1}}{n_1!}\frac{\Gamma(\tfrac{d}{2}-a_{34}-n_1)\Gamma(\tfrac{d}{2}+n_1-a_1)}{\Gamma(a_1-n_1)}\int [ds] \Gamma(-s)\Gamma(\tfrac{d}{2}-a_{23}-s)\Gamma(a_{15}-\tfrac{d}{2}-n_1-s) \crcr 
    & \qquad \frac{\Gamma(\tfrac{d}{2}-a_5+s)\Gamma(a_3+n_1+s)\Gamma(a_{234}-\tfrac{d}{2}+n_1+s)}{\Gamma(a_5-s)\Gamma(d-a_{15}+n_1+s)} \, , \crcr
    I_b&= \sum_{n_1\geq 0}\frac{(-1)^{n_1}}{n_1!}\frac{\Gamma(a_{34}-\tfrac{d}{2}-n_1)\Gamma(d+n_1-a_{134})}{\Gamma(a_{134}-\tfrac{d}{2}-n_1)}\int [ds] \Gamma(-s)\Gamma(\tfrac{d}{2}-a_{23}-s)\Gamma(a_{1345}-d-n_1-s) \crcr 
    & \qquad \frac{\Gamma(\tfrac{d}{2}-a_5+s)\Gamma(a_2+n_1+s)\Gamma(\tfrac{d}{2}+n_1-a_4+s)}{\Gamma(a_5-s)\Gamma(\tfrac{3d}{2}-a_{1345}+n_1+s)} \, , \crcr
    I_c&= \sum_{n_1\geq 0}\frac{(-1)^{n_1}}{n_1!}\frac{\Gamma(a_{12345}-d+n_1)\Gamma(a_{135}-\tfrac{d}{2}+n_1)}{\Gamma(\tfrac{d}{2}+n_1)}\int [ds] \Gamma(-s)\Gamma(\tfrac{d}{2}-a_{23}-s)\Gamma(a_{5}+n_1-s) \crcr 
    & \qquad \frac{\Gamma(\tfrac{d}{2}-a_5+s)\Gamma(d-a_{1345}-n_1+s)\Gamma(\tfrac{d}{2}-n_1-a_{15}+s)}{\Gamma(a_5-s)\Gamma(\tfrac{d}{2}-a_{5}-n_1+s)} \, , 
\end{align}
where we used the notation $a_{123}=a_1+a_2+a_3$ and so on. 

We now close the contour for $s$. For $I_a$, we have three families of poles located at $s=n_2$, $s=d/2-a_{23}+n_2$ and $s=a_{15}-d/2+n_2$ for $n_2\geq 0$. 
We thus have $I_a=I_{a,1}+I_{a,2}+I_{a,3}$ with:
\begin{align}
    I_{a,1}=\sum_{n_1,n_2\geq 0}\frac{(-1)^{n_1+n_2}}{n_1!n_2!}&\frac{\Gamma(\tfrac{d}{2}-a_{34}-n_1)\Gamma(\tfrac{d}{2}-a_1+n_1)\Gamma(\tfrac{d}{2}-a_{23}-n_2)\Gamma(\tfrac{d}{2}-a_5+n_2)}{\Gamma(a_1-n_1)\Gamma(a_5-n_2)}  \crcr 
    &\times  \frac{\Gamma(a_{15}-\tfrac{d}{2}-n_1-n_2)\Gamma(a_3+n_1+n_2)\Gamma(a_{234}-\tfrac{d}{2}+n_1+n_2)}{\Gamma(d-a_{15}+n_1+n_2)} \, , \crcr
    I_{a,2}=\sum_{n_1,n_2\geq 0}\frac{(-1)^{n_1+n_2}}{n_1!n_2!}&\frac{\Gamma(\tfrac{d}{2}-a_{34}-n_1)\Gamma(\tfrac{d}{2}-a_1+n_1)\Gamma(a_{23}-\tfrac{d}{2}-n_2)\Gamma(d-a_{235}+n_2)}{\Gamma(a_1-n_1)\Gamma(a_{235}-\tfrac{d}{2}-n_2)}  \crcr 
    &\times  \frac{\Gamma(a_{1235}-d-n_1-n_2)\Gamma(a_4+n_1+n_2)\Gamma(\tfrac{d}{2}-a_2+n_1+n_2)}{\Gamma(\tfrac{3d}{2}-a_{1235}+n_1+n_2)} \, , \crcr
    I_{a,3}=\sum_{n_1,n_2\geq 0}\frac{(-1)^{n_2}}{n_1!(n_1+n_2)!}&\frac{\Gamma(\tfrac{d}{2}-a_{34}-n_1)\Gamma(\tfrac{d}{2}-a_1+n_1)\Gamma(\tfrac{d}{2}-a_{15}-n_2)\Gamma(d-a_{1235}-n_2)\Gamma(a_1+n_2)}{\Gamma(a_1-n_1)\Gamma(\tfrac{d}{2}-a_1-n_2)}  \crcr 
    &\times  \frac{\Gamma(a_{12345}-d+n_1+n_2)\Gamma(a_{135}-\tfrac{d}{2}+n_1+n_2)}{\Gamma(\tfrac{d}{2}+n_1+n_2)} \, . 
\end{align}
Notice that if $a_5$ and $a_1$ are positive integers, $I_{a,1}$ reduces to the sum over $n_1<a_1$ and $n_2<a_5$ and $I_{a,2}$ and $I_{a,3}$ reduce to the sum over $n_1<a_1$.

For $I_b$, we have again three families of poles located at $s=n_2$, $s=d/2-a_{23}+n_2$ and $s=a_{1345}-d+n_2$ for $n_2\geq 0$. We thus have $I_b=I_{b,1}+I_{b,2}+I_{b,3}$ with:

\begin{align}
    I_{b,1}=\sum_{n_1,n_2\geq 0}\frac{(-1)^{n_1+n_2}}{n_1!n_2!}&\frac{\Gamma(a_{34}-\tfrac{d}{2}-n_1)\Gamma(d+n_1-a_{134})\Gamma(\tfrac{d}{2}-a_{23}-n_2)\Gamma(\tfrac{d}{2}-a_5+n_2)}{\Gamma(a_{134}-\tfrac{d}{2}-n_1)\Gamma(a_5-n_2)} \crcr
    & \times \frac{\Gamma(a_{1345}-d-n_1-n_2)\Gamma(a_2+n_1+n_2)\Gamma(\tfrac{d}{2}-a_4+n_1+n_2)}{\Gamma(\tfrac{3d}{2}-a_{1345}+n_1+n_2)} \, , \crcr 
     I_{b,2}=\sum_{n_1,n_2\geq 0}\frac{(-1)^{n_1+n_2}}{n_1!n_2!}&\frac{\Gamma(a_{34}-\tfrac{d}{2}-n_1)\Gamma(d+n_1-a_{134})\Gamma(a_{23}-\tfrac{d}{2}-n_2)\Gamma(d-a_{235}+n_2)}{\Gamma(a_{134}-\tfrac{d}{2}-n_1)\Gamma(a_{235}-\tfrac{d}{2}-n_2)} \crcr
    & \times \frac{\Gamma(a_{12345}+a_3-\tfrac{3d}{2}-n_1-n_2)\Gamma(\tfrac{d}{2}-a_3+n_1+n_2)\Gamma(d-a_{234}+n_1+n_2)}{\Gamma(2d-a_{12345}-a_3+n_1+n_2)} \, , \crcr 
     I_{b,3}=\sum_{n_1,n_2\geq 0}\frac{(-1)^{n_2}}{n_1!(n_1+n_2)!}&\frac{\Gamma(a_{34}-\tfrac{d}{2}-n_1)\Gamma(d+n_1-a_{134})\Gamma(d-a_{1345}-n_2)\Gamma(\tfrac{3d}{2}-a_{12345}-a_3-n_2)}{\Gamma(a_{134}-\tfrac{d}{2}-n_1)\Gamma(d-a_{134}-n_2)} \crcr
    & \times \frac{\Gamma(a_{134}-\tfrac{d}{2}+n_2)\Gamma(a_{135}-\tfrac{d}{2}+n_1+n_2)\Gamma(a_{12345}-d+n_1+n_2)}{\Gamma(\tfrac{d}{2}+n_1+n_2)} \, .
\end{align}
Notice that if $a_5$ is a positive integer $I_{b,1}$ reduces to the sum over $n_2<a_5$. 

For $I_c$ we have four families of poles located at $s=n_2$, $s=d/2-a_{23}+n_2$ for $n_2\geq 0$ and $s=a_{15}-d/2+n_1-n_2$, $s=a_{1345}-d+n_1-n_2$ for $0\leq n_2 \leq n_1$. We then have $I_c=I_{c,1}+I_{c,2}+I_{c,3}+I_{c,4}$ with:

\begin{align}
    I_{c,1}=\sum_{n_1,n_2\geq 0}\frac{(-1)^{n_1+n_2}}{n_1!n_2!}&\frac{\Gamma(a_{12345}-d+n_1)\Gamma(a_{135}-\tfrac{d}{2}+n_1)\Gamma(\tfrac{d}{2}-a_{23}-n_2)\Gamma(\tfrac{d}{2}-a_5+n_2)}{\Gamma(\tfrac{d}{2}+n_1)\Gamma(a_5-n_2)} \crcr
    & \times \frac{\Gamma(a_5+n_1-n_2)\Gamma(\tfrac{d}{2}-a_{15}-n_1+n_2)\Gamma(d-a_{1345}-n_1+n_2)}{\Gamma(\tfrac{d}{2}-a_5-n_1+n_2)} \, , \crcr
    I_{c,2}=\sum_{n_1,n_2\geq 0}\frac{(-1)^{n_1+n_2}}{n_1!n_2!}&\frac{\Gamma(a_{12345}-d+n_1)\Gamma(a_{135}-\tfrac{d}{2}+n_1)\Gamma(a_{23}-\tfrac{d}{2}-n_2)\Gamma(d-a_{235}+n_2)}{\Gamma(\tfrac{d}{2}+n_1)\Gamma(a_{235}-\tfrac{d}{2}-n_2)} \crcr
    & \times \frac{\Gamma(a_{235}-\tfrac{d}{2}+n_1-n_2)\Gamma(\tfrac{3d}{2}-a_{12345}-a_3-n_1+n_2)\Gamma(d-a_{1235}-n_1+n_2)}{\Gamma(d-a_{235}-n_1+n_2)} \, , \crcr
    I_{c,3}=\sum_{n_1,n_2\geq 0}\frac{(-1)^{n_1+1}}{(n_1+n_2)!n_2!}&\frac{\Gamma(a_{12345}-d+n_1+n_2)\Gamma(a_{135}-\tfrac{d}{2}+n_1+n_2)\Gamma(\tfrac{d}{2}-a_{34}-n_2)\Gamma(\tfrac{d}{2}-a_1+n_2)}{\Gamma(\tfrac{d}{2}+n_1+n_2)\Gamma(a_1-n_2)} \crcr
    & \times \frac{\Gamma(a_1+n_1)\Gamma(\tfrac{d}{2}-a_{15}-n_1)\Gamma(d-a_{1235}-n_1)}{\Gamma(\tfrac{d}{2}-a_1-n_1)} \, , \crcr
    I_{c,4}=\sum_{n_1,n_2\geq 0}\frac{(-1)^{n_1+1}}{(n_1+n_2)!n_2!}&\frac{\Gamma(a_{12345}-d+n_1+n_2)\Gamma(a_{135}-\tfrac{d}{2}+n_1+n_2)\Gamma(a_{34}-\tfrac{d}{2}-n_2)\Gamma(d-a_{134}+n_2)}{\Gamma(\tfrac{d}{2}+n_1+n_2)\Gamma(a_{134}-\tfrac{d}{2}-n_2)} \crcr
    & \times \frac{\Gamma(a_{134}-\tfrac{d}{2}+n_1)\Gamma(\tfrac{3d}{2}-a_{12345}-a_3-n_1)\Gamma(d-a_{1345}-n_1)}{\Gamma(d-a_{134}-n_1)} \, ,
\end{align}
where for $I_{c,3}$ and $I_{c,4}$ we have inverted the sums and perform the change of variable $n_1\rightarrow n_1+n_2$ so that both sums run from $0$ to $\infty$. 
Notice that if $a_1$ is a positive integer $I_{c,3}$ reduces to the sum over $n_2<a_1$. 

To facilitate the analysis of the divergences when specifying values for the $a_i$ and $d$, we separate $I_{c,1}$ and $I_{c,2}$ into the $n_2\leq n_1$ and $n_2> n_1$ parts and perform a change of variable in the sums so that they both run from $0$ to $\infty$:

\begin{align}
    I_{c,1}&=\sum_{n_1,n_2\geq 0}\frac{(-1)^{n_1}}{(n_1+n_2)!n_2!}\frac{\Gamma(a_{12345}-d+n_1+n_2)\Gamma(a_{135}-\tfrac{d}{2}+n_1+n_2)\Gamma(\tfrac{d}{2}-a_{23}-n_2)}{\Gamma(\tfrac{d}{2}+n_1+n_2)\Gamma(a_5-n_2)} \crcr
    & \qquad \qquad \times \frac{\Gamma(\tfrac{d}{2}-a_5+n_2)\Gamma(a_5+n_1)\Gamma(\tfrac{d}{2}-a_{15}-n_1)\Gamma(d-a_{1345}-n_1)}{\Gamma(\tfrac{d}{2}-a_5-n_1)} \crcr
    +&\sum_{n_1,n_2\geq 0}\frac{(-1)^{1+n_2}}{(1+n_1+n_2)!n_1!}\frac{\Gamma(a_{12345}-d+n_1)\Gamma(a_{135}-\tfrac{d}{2}+n_1)\Gamma(\tfrac{d}{2}-a_{23}-1-n_1-n_2)}{\Gamma(\tfrac{d}{2}+n_1)\Gamma(a_5-1-n_1-n_2)} \crcr
    & \times \frac{\Gamma(\tfrac{d}{2}-a_5+1+n_1+n_2)\Gamma(a_5-1-n_2)\Gamma(\tfrac{d}{2}-a_{15}+1+n_2)\Gamma(d-a_{1345}+1+n_2)}{\Gamma(\tfrac{d}{2}-a_5+1+n_2)}  \, , 
\end{align}

\begin{align}
    I_{c,2}&=\sum_{n_1,n_2\geq 0}\frac{(-1)^{n_1}}{(n_1+n_2)!n_2!}\frac{\Gamma(a_{12345}-d+n_1+n_2)\Gamma(a_{135}-\tfrac{d}{2}+n_1+n_2)\Gamma(a_{23}-\tfrac{d}{2}-n_2)}{\Gamma(\tfrac{d}{2}+n_1+n_2)\Gamma(a_{235}-\tfrac{d}{2}-n_2)} \crcr
    & \qquad \qquad\times \frac{\Gamma(d-a_{235}+n_2)\Gamma(a_{235}-\tfrac{d}{2}+n_1)\Gamma(d-a_{1235}-n_1)\Gamma(\tfrac{3d}{2}-a_{12345}-a_3-n_1)}{\Gamma(d-a_{235}-n_1)} \crcr
    +&\sum_{n_1,n_2\geq 0}\frac{(-1)^{1+n_2}}{(1+n_1+n_2)!n_1!}\frac{\Gamma(a_{12345}-d+n_1)\Gamma(a_{135}-\tfrac{d}{2}+n_1)\Gamma(a_{23}-\tfrac{d}{2}-1-n_1-n_2)}{\Gamma(\tfrac{d}{2}+n_1)\Gamma(a_{235}-\tfrac{d}{2}-1-n_1-n_2)} \crcr
    & \qquad \qquad\times \frac{\Gamma(d-a_{235}+1+n_1+n_2)\Gamma(a_{235}-\tfrac{d}{2}-1-n_2)}{\Gamma(d-a_{235}+1+n_2)}  \crcr
    & \qquad \qquad  \times \Gamma(\tfrac{3d}{2}-a_{12345}-a_3+1+n_2)\Gamma(d-a_{1235}+1+n_2) \, .
\end{align}
Notice that if $a_5$ is a positive integer the first sum in $I_{c,1}$ reduces to the sum over $n_2<a_5$. 

We thus wrote the generic kite integral as a double-sum of ratios of Gamma functions. 

\subsubsection{Check for I(1,1,1,1,1) and $d=4-\epsilon$}

For $a_1=a_2=a_3=a_4=a_5=1$ and $d=4-\epsilon$, it is known from the integration by part method that $I(1,1,1,1,1)= \frac{p^{-2-2\epsilon}}{(4\pi)^4}6\zeta(3)$. As a sanity check, we aim to benchmark our generic result with this exact particular case.
To this end we substitute the values of the $a_i$ and of $d$ in every term of the double-sum, expand in $\epsilon$ to order $\mathcal{O}(\epsilon^0)$ and take the sum. Even though the final result is of order $\mathcal{O}(\epsilon^0)$, we cannot simply take $\epsilon=0$ in each term as the individual terms display divergences in $1/\epsilon$ that cancel only overall. 

As we pointed out, some terms in the double sum are zero when $n_1$ or $n_2$ are non-zero. We thus separate our sum into four regions : $n_1=n_2=0$, $n_1=0, n_2\geq 1$, $n_1\geq 1, n_2=0$ and $n_1,n_2 \geq 1$ and we denote $I(1,1,1,1,1)=\frac{p^{-2-2\epsilon}}{(4\pi)^4}\left(I_{0,0}+I_{0,1}+I_{1,0}+I_{1,1}\right)$. 

\paragraph{$n_1=n_2=0$}

The terms $n_1=n_2=0$ are non-zero for every family of poles. 
Grouping them together and expanding in $\epsilon$, we have $I_{0,0}=4+\frac{\pi^2}{3}$. 

\paragraph{$n_1=0, \, n_2\geq 1$}

In this case the terms coming from $I_{a,1}, I_{b,1}$, from the first sum in $I_{c,1}$ and from $I_{c,3}$ do not contribute. 
Grouping the other terms and expanding in $\epsilon$, we obtain:
\begin{equation}
    I_{0,1}=\sum_{n_2\geq 1} \frac{1+2n_2+2n_2^3}{n_2^3(1+n_2)^3}=4(\zeta(3)-1) \, .
\end{equation}

\paragraph{$n_1 \geq 1, \, n_2=0$}

In this case the terms coming from $I_{a,1},I_{a,2}$ and $I_{a,3}$ do not contribute. 
Grouping the other terms and expanding in $\epsilon$, we obtain:
\begin{equation}
    I_{1,0}=-\sum_{n_1\geq 1} \frac{3+\pi^2+(\pi^2-3)n_1(2+n_1)}{3n_1(1+n_1)^3}=2\zeta(3)-5 \, .
\end{equation}

\paragraph{$n_1 \geq 1, \, n_2 \geq 1$}

In this case the terms coming from $I_{a,1},I_{a,2},I_{a,3},I_{b,1}, I_{c,3}$ and from the first sum in $I_{c,1}$ do not contribute. 
Grouping the other terms and expanding in $\epsilon$, we obtain:
\begin{equation}
\begin{split}
  I_{1,1}& =-\sum_{n_1\geq 1, n_2\geq 1} \frac{1}{n_1(n_1+1)}\left( \frac{1}{n_2^2}+\frac{1}{(1+n_2)^2}-\frac{2}{(1+n_1+n_2)^2}\right) \crcr
  & =\sum_{n_1\geq 1}\frac{\pi^2-3-6\psi_1(n_1+2)}{3n_1(n_1+1)}=\frac{15-\pi^2}{3} \, ,
\end{split}
\end{equation}
and grouping everything together we reproduce $I(1,1,1,1,1)= \frac{p^{-2-2\epsilon}}{(4\pi)^4}6\zeta(3)$. 

\subsubsection{Computation of $K_0$}

In order to compute $K_0$, we use the previous decomposition in double-sums with $a_1=a_2=a_3=a_4=a_5=\zeta=\frac{d+\epsilon}{4}$.
In this case all the terms will contribute and every sum will have to be evaluated for $n_1\geq 0,n_2\geq 0$. Before summing, we expand the summand in $\epsilon$. As expected, all divergences cancel and we obtain an expression of order $\mathcal{O}(\epsilon^0)$. As the resulting expression is too lengthy to write here, we provide it in an ancillary file. We can evaluate the sum numerically with good precision and obtain:
\begin{align}
    \alpha_{I_4}&=30.1026152(6) \, , \qquad \text{at } d=3 \, ,\crcr
    \alpha_{I_4}&= 76.62828703(7)\, , \qquad \text{at } d=2 \, , \crcr
    \alpha_{I_4}&= 330.008919215(21)\, ,\qquad \text{at } d=1 \, . 
\end{align}

We note that, $K_0$ being finite, we could also have taken $\epsilon=0$ from the beginning in the integral expression 
in Eq.~\eqref{eq:kite}. This leads to poles of order $4$, whose residues can be computed using the package MBsums. However, the resulting double sum is much less amenable to numerical evaluation and using this alternative method we were not able to reach the same numerical precision.  

\subsection{Computation with Hepp sectors}

As a cross-check of our result, we now evaluate $I_4$ using the same infrared regularization used for the other graphs in \cite{Benedetti:2020rrq}, that is, we set the external momenta to zero and use the regularized covariance $C_{\mu}(p)=\tfrac{1}{\left(p^2+\mu^2 \right)^{\zeta}}$. We can then write $I_4$ in terms of Schwinger parameters:
\begin{equation}
    I_4=\frac{1}{(4\pi)^{3d/2}\Gamma(d/4)^6}\int_0^{\infty} da_1 \dots dc_2\frac{(a_1a_2b_1b_2c_1c_2)^{\zeta-1}e^{-(a_1+a_2+b_1+b_2+c_1+c_2)}}{\mathcal{S}^{d/2}} \, ,
\end{equation}
with 
\begin{align}
\mathcal{S}=&a_1a_2(b_1+b_2+c_1+c_2)+b_1b_2(a_1+a_2+c_1+c_2)+c_1c_2(a_1+a_2+b_1+b_2)\crcr
&+a_1(b_2c_1+b_1c_2)+a_2(b_1c_1+b_2c_2) \, .
\end{align}

Next, we extract the divergence in $1/\epsilon$ by introducing Hepp sectors. The denominator $\mathcal{S}$ is symmetric by exchange of the pairs $(a_1,a_2),(b_1,b_2),(c_1,c_2)$ and by simultaneous exchange of elements of two of the pairs. Therefore, instead of $6!=720$ sectors we only have $\frac{6!}{3! 2^2}=30$ sectors to consider. 

For example, let us choose the sector $a_1\geq a_2\geq b_1 \geq b_2\geq c_1\geq c_2$ and perform the following change of variable:
\begin{align}
    & a_2=t_1 a_1 \, ,\crcr
    & b_1= t_2 t_1 a_1 \, , \, b_2= t_3 t_2 t_1 a_1 \, ,\crcr
    & c_1= t_4 t_3 t_2 t_1 a_1 \, , \, c_2= t_5 t_4 t_3 t_2 t_1 a_1 \, .
\end{align}
The integral for this sector is:
\be
\begin{split}
    I_{4,1}=\frac{48}{(4\pi)^{\tfrac{3d}{2}}\Gamma(\tfrac{d}{4})^6}\int_0^{\infty} da_1 \int_0^1 \prod_{i=1}^5 dt_i &\frac{a_1^{5-3d/2} t_1^4 t_2^3t_3^2 t_4 (a_1^6 t_1^5t_2^4t_3^3t_4^2t_5)^{\zeta-1}}{(t_1^2t_2 \mathcal{S}_1)^{d/2}} \\
    & \quad \times e^{-a_1(1+t_1+t_1t_2+t_1t_2t_3t+t_1t_2t_3t_4+t_1t_2t_3t_4t_5)} \, ,
\end{split}
\ee
with:
\begin{align}
    \mathcal{S}_1=1 + t_3 \big[1 &+ t_4 + t_2 (1 + t_1 + (t_1 + t_3 + t_1 t_2 t_3) t_4) \crcr 
    &+ 
    t_4 (1 + t_2 + t_2 t_3 t_4 + 
       t_1 t_2 t_3 (1 + t_4 + t_2 (1 + t_4 + t_3 t_4))) t_5\big] \, .
\end{align}
We now integrate $a_1$, which factors a $\Gamma(3\epsilon/2)$. Since the full integral is of order $\mathcal{O}(\epsilon^{-1})$, we can then take $\epsilon=0$ in the remaining integral and obtain:
\begin{align}
    &I_{4,1}=\frac{32 }{\epsilon(4\pi)^{\tfrac{3d}{2}}\Gamma(\tfrac{d}{4})^6} \int_0^1 \prod_{i=1}^5 dt_i  \frac{\left(t_1t_5 \right)^{d/4-1}\left(t_2t_4 \right)^{d/2-1}t_3^{3d/4-1}}{ \mathcal{S}_1^{d/2}} +\mathcal{O}(\epsilon^0)\, .
\end{align}
The remaining integral is convergent\footnote{As the $t_i$ are positive $\mathcal{S}_1\geq 1$ and $0\leq \int_0^1 \prod_{i=1}^5 dt_i  \frac{\left(t_1t_5 \right)^{d/4-1}\left(t_2t_4 \right)^{d/2-1}t_3^{3d/4-1}}{ \mathcal{S}_1^{d/2}}\leq \int_0^1 \prod_{i=1}^5 dt_i  \left(t_1t_5 \right)^{d/4-1}\left(t_2t_4 \right)^{d/2-1}t_3^{3d/4-1}$, yielding an upper bound $\frac{4096}{d^5}$.} and can then be evaluated numerically. 

Repeating the same steps for all non-equivalent sectors we obtain the same values for $\alpha_{I_4}$ as with the previous method up to the eighth digit. 

This serves as a cross-check of our results and provides enough significant digits for our analysis below. If we wish to improve precision, we can sum more terms in the first method or improve the precision of the numerical integration in the Hepp sectors method by performing integrations by parts and working with a higher precision in the Mathematica function NIntegrate.

\section{Applications}
\label{sec:appl}

We recap here the numerical results of \cite{Benedetti:2020rrq} corrected with the new value of $\alpha_{I_4}$. 

\subsection{The long-range Ising model}
\label{sec:ising model}

For the long-range Ising model in integer dimensions, the inverse of the correlation length exponent $\nu^{-1}= - \, \partial_{\rt}\beta^{(2)}(\gt_{\star})$ evaluates to:
\begin{align}
\nu^{-1} \, & = \, 1.5 + 0.1667\, \epsilon - 0.1812\, \epsilon^2 +  0.3536\, \epsilon^3 \,, & \text{at } \; d=3\,, \crcr
\nu^{-1} \, & = \,  1 + 0.1667\, \epsilon - 0.3081\, \epsilon^2 +  0.8749\, \epsilon^3 \,, & \text{at } \; d=2\,, \crcr
\nu^{-1} \, & = \,  0.5 + 0.1667\, \epsilon- 0.6571\, \epsilon^2 + 3.720\, \epsilon^3 \,, & \text{at } \; d=1\,.
\end{align}

In Table~\ref{tab:critexp} we compare the numerical values of $\nu$ in dimension $d=1$ for different values of $\epsilon$ obtained from our loop expansion with those from numerical simulations \cite{Glumac:1989,Luijten:1997-thesis,uzelac2001critical,tomita2009monte}. In Table~\ref{tab:critexp-omega} we provide a similar comparison for the susceptibility exponent $\omega=\partial_{\gt}\beta^{(4)}(\gt_{\star})$ in dimension $d=1$ with available numerical simulations \cite{Luijten:1997-thesis}. In Table \ref{tab:critexp2} we compare the numerical values of $\nu$ in dimension $d=2$ at different values of $\epsilon$ obtained with our method with numerical simulations \cite{Angelini:2014}. The value $\epsilon=1.5$ is the maximum value of $\epsilon$ we consider because it corresponds to the transition between long-range and short-range behavior happening at $2\zeta=2-\eta_{SR}$. We find a value of $\n$ consistent with the exact result $\n_{SR}=1$ in $d=2$. 

\begin{table}[htb]
\begin{center}
\begin{tabular}{|c|c||c|c|c|c|c|c|c|c|}\hline
$\epsilon$ &  $2\zeta$  &  mean-field      &     three-loop   &    PB $[2/1]$ &   Ref.~\cite{Glumac:1989}        &     Ref.~\cite{Luijten:1997-thesis}  & Ref.~\cite{uzelac2001critical} & Ref.~\cite{tomita2009monte}                  \\ \hline
0.2 & 0.6 &1.6667    &     1.8471    &  1.9247(32)   &     2.16        &     1.98      & 2.00(16) & 1.80(21)    \\
0.4 & 0.7 & 1.4286   &      1.1203        &  1.896(13)  &    2.123       &      2.01 & 1.96(15) & 1.83(17)               \\
0.6 & 0.8 & 1.25   &    -0.98220  &  1.885(26)  &     2.208        &       2.17 & 2.13(18) & 1.89(5)              \\
\hline
\end{tabular}\end{center}
\caption{The critical exponent $\nu$ for the long-range Ising model at $d=1$, as computed in mean-field, $\nu=(2\z)^{-1}$, by the naive three-loop series, and by a Pad\'e-Borel (PB) summation with $[2/1]$ approximant. The error in the latter is estimated by the difference with the PB summation of the two-loop series with $[1/1]$ approximant. The last four columns report numerical results from the literature for comparison.
}
\label{tab:critexp}
\end{table}
\begin{table}[htb]
\begin{center}
\begin{tabular}{|c|c||c|c|c|c|}\hline
$\epsilon$ &  $2\zeta$ &   one-loop         &     three-loop    &    PB $[2/1]$       &     Ref.~\cite{Luijten:1997-thesis}                    \\ \hline
0.2 & 0.6 & 0.2          &     0.43    &  0.147(25)          &     0.15         \\
0.4 & 0.7 & 0.4      &     2.86      &  0.27(8)         &      0.23                 \\
0.6 & 0.8 & 0.6       &    9.60   &  0.38(16)       &       0.25              \\
\hline
\end{tabular}\end{center}
\caption{The critical exponent $\omega$ for the long-range Ising model at $d=1$, as computed by one- and three-loop truncations and by a Pad\'e-Borel summation of the three-loop series with $[2/1]$ approximant, with errors estimated as before. The last column reports numerical results from Ref.~\cite{Luijten:1997-thesis}  for comparison.
}
\label{tab:critexp-omega}
\end{table}

\begin{table}[htb]
\begin{center}
\begin{tabular}{|c|c||c|c|c|c|}\hline
$\epsilon$ &  $2\zeta$  & mean-field &     three-loop    &   PB $[2/1]$ &   Ref.~\cite{Angelini:2014}                         \\ \hline
0.4 & 1.2 &0.8333 &0.9242  & 0.9607(8) &   0.977(34)     \\
0.61812 & 1.30906 & 0.7640& 0.7933 & 0.9499(21) & 0.986(33)     \\
1.2 & 1.6 & 0.625& -0.4136 & 0.935(7) &  1.004(34)   \\
1.5 & 1.75 &0.5714 & -1.8092  & 0.931(11) &  1.02 (12) \\
\hline
\end{tabular}\end{center}
\caption{The critical exponent $\nu$ for the long-range Ising model at $d=2$, as computed in mean-field, $\nu=(2\z)^{-1}$, by the naive three-loop series, and by a Pad\'e-Borel (PB) summation with $[2/1]$ approximant (with error estimated by the difference with the PB summation of the two-loop series with $[1/1]$ approximant). The last column reports numerical results from the literature for comparison.}
\label{tab:critexp2}
\end{table}

\paragraph{On the conjectured relation between short-range and long-range Ising models.}
The peculiar value $\epsilon=0.61812$ was considered in \cite{Angelini:2014} (and hence in our  Table \ref{tab:critexp2}) in order to test a conjecture relating the long-range Ising model at given dimension $d$, to the short-range Ising model at a different dimension $d_{SR}$ \cite{Banos:2012,Angelini:2014,Defenu:2014}. According to the conjecture, the effective dimension or effective $\z$ should be found from the relation:
 \be \label{eq:LR-SR-1}
 \f{2\zeta}{d}= \f{2-\eta_{SR}(d_{SR})}{d_{SR}}\,,
 \ee
 and one should then also see other relations among the critical exponents, such as: 
 \be \label{eq:LR-SR-2}
 d\,\nu(\z,d)=d_{SR}\, \nu_{SR}(d_{SR})\,, \quad \omega(\z,d)/d=\omega_{SR}(d_{SR})/d_{SR}\,.
 \ee
Therefore, the short-range model at $d_{SR}=3$ should be related to the long-range model at $d=2$ if one takes  $2\zeta=1.30906$, or equivalently $\epsilon=0.61812$.
In \cite{Benedetti:2020rrq}, we showed that these relations were at best approximate and qualitative. 
Comparing the short-range results from the literature \cite{El-Showk:2014dwa}, $3\nu_{SR}(3)\simeq 1.88997(15)$, with our resummed result $2\nu_{LR}\simeq 1.898(4)$, we find that the values are approximately compatible with equation \eqref{eq:LR-SR-2}, within the given errors. 
From the Pad\'e-Borel summation of our three-loop result  at $d=2$ we find that $\omega/2\simeq 0.21(5)$, which is compatible, within the given errors, with the value $\omega_{SR}(3)/3 \simeq 0.2767(6)$ obtained again from \cite{El-Showk:2014dwa}.
We thus conclude that as an effect of our correction to $I_4$, we find an improvement of the numerical matching between the short-range Ising model in $d_{SR}=3$ and the two-dimensional long-range Ising model with $\z$ satisfying \eqref{eq:LR-SR-1}.

\subsection{The long-range $O(N)$ vector model}
\label{sec:vector model}

At low $N$ and integer $d$, the series expansion of the inverse of the correlation length exponent is:
\begin{align}
\nu^{-1}&=0.5+0.1\epsilon-0.8043\epsilon^2+3.766\epsilon^3 \,, \quad (d=1, N=2 ) \,, \crcr
\nu^{-1}&=1+0.1\epsilon-0.3771\epsilon^2+0.8880\epsilon^3 \,, \quad ( d=2, N=2 ) \,, \crcr
\nu^{-1}&=1.5+0.1\epsilon-0.2218\epsilon^2+0.3608\epsilon^3 \,, \quad ( d=3 , N=2 ) \,, \crcr
\nu^{-1}&=0.5+0.04545\epsilon-0.9109\epsilon^2+3.705\epsilon^3 \,, \quad ( d=1, N=3 ) \,, \crcr
\nu^{-1}&=1+0.04545\epsilon-0.4270\epsilon^2+0.8740\epsilon^3 \,, \quad ( d=2, N=3 ) \,, \crcr
\nu^{-1}&=1.5+0.04545\epsilon-0.2512\epsilon^2+0.3556\epsilon^3 \,, \quad ( d=3, N=3 ) \,.
\end{align}

In Table \ref{tab:critvect22} we report the numerical values of $\nu$ and $\omega$ in two dimensions at $N=2$ and in three dimensions at $N=2,3$ for different values of $\epsilon$. 

\begin{table}[htb]
\begin{center}
\begin{tabular}{|c|c|c||c|c|c| |c|c|c| }\hline
\multirow{2}{*}{$(d,N)$ } & \multirow{2}{*} {$\epsilon$} &  \multirow{2}{*}{$2\zeta$} &  \multicolumn{3}{c||}{$\nu$}  
& 
 \multicolumn{2}{c|}{ $\omega$ }\\
& & & mean-field  &     three-loop    &    PB $[2/1]$  & three-loop &  PB $[2/1]$\\ \hline
\multirow{3}{*}{(2,2)}
& 0.2 & 1.1 &0.9091     &     0.9878   &  0.9906(29)  & 0.2088 & 0.164(12)\\
& 0.4 & 1.2 & 0.8333    &   1.019      &  0.993(12) & 0.6829 & 0.299(47)\\
& 0.6 & 1.3 & 0.7692  &    1.067   &  1.002(25)  & 1.660 & 0.42(10)\\
\hline
\multirow{3}{*}{(3,2)} 
& 0.2 & 1.6 &0.625  &     0.6605   &  0.6609(6)  & 0.1872 & 0.174(6) \\
& 0.4 & 1.7 & 0.5882 &      0.6762      &  0.6590(26)  & 0.4406 & 0.323(30)\\
& 0.6 & 1.8 & 0.5556  &    0.7052   &  0.659(6)   &  0.8446 & 0.46(7)\\
\hline
\multirow{3}{*}{(3,3)} 
& 0.2 & 1.6 &0.625    &     0.9909   &  0.6662(16)   & 0.2344 & 0.175(6)\\
& 0.4 & 1.7 & 0.5882 &      0.6762      &  0.670(7)  & 0.4406 & 0.324(28)\\
& 0.6 & 1.8 & 0.5556 &    0.7052   &  0.677(14)   & 0.8446  &  0.46(6) \\
\hline
\end{tabular}\end{center}
\caption{The critical exponents $\nu$ and $\omega$ for the long-range vector model at various $d$ and $N$ computed in mean-field, $\nu=(2\z)^{-1}$, by the naive three-loop series, and by a Pad\'e-Borel (PB) summation with $[2/1]$ approximant (with error estimated by the difference with the PB summation of the two-loop series with $[1/1]$ approximant).}
\label{tab:critvect22}
\end{table}

\subsection{The long-range cubic model}
\label{sec:cubic}

The next model we consider is obtained by breaking explicitly the $O(N)$ symmetry with an interaction of the form $\sum_{\mba} \phi_{\mba}^4$: this is the (hyper)-cubic model with symmetry $(\mathbb{Z}_2)^N \rtimes S_N$.
We give below numerical values of critical exponents with the corrected value of $I_4$.

There are three non-trivial fixed points: the Ising fixed point with $g_d=0$, the Heisenberg fixed point with $g_c=0$ and the cubic fixed point with both couplings non-zero.
The stability of the Heisenberg and cubic fixed points are related. Indeed, there exists a critical value $N=N_c$ for which the cubic and Heisenberg fixed points collapse and exchange stability, as in the short-range model \cite{Kleinert:2001ax}. The Heisenberg fixed point is stable for $N<N_c$ and the cubic fixed point is stable for  $N>N_c$. 
Table \ref{tab:nc} gives values of $N_c$ at $d=3$ for different values of $\epsilon$ obtained at different loop-orders and with the Pad\'e-Borel summation method. 
We can also study the numerics of the critical exponents $\nu_C$ and  $\nu_H$ of the cubic and Heisenberg fixed points at $N=3$ for different values of $\epsilon$. The exponent $\nu_H$ at $d=3,N=3$ is identical with the exponent $\nu$ reported in table \ref{tab:critvect22}. For comparison, the corresponding $\nu_C$ is displayed in table \ref{tab:nucubic}.

\begin{table}[htb]
\begin{center}
\begin{tabular}{|c||c|c|c|}\hline
$\epsilon$ &   one-loop          &     three-loop    &    PB $[1/1]$  \\ \hline
0.2 & 4 &        3.6322   &  3.528(33)  \\
 0.4 & 4 &      3.8334      &  3.24(8)                 \\
 0.6 & 4 &      4.6038  &  3.03(12)   \\
\hline
\end{tabular}\end{center}
\caption{The critical value $N_c$ for the long-range cubic model at $d=3$, as computed by a one-loop and three-loop truncation and by a Pad\'e-Borel summation of the three-loop series with $[1/1]$ approximant (the error is estimated by the difference with the PB summation of the two-loop series with $[0/1]$ approximant).}
\label{tab:nc}
\end{table}

\begin{table}[h]
\begin{center}
\begin{tabular}{|c|c||c|c|c|}\hline
$\epsilon$& $2\zeta$ &  mean-field &   three-loop     &   PB $[2/1]$      \\ \hline
0.2 & 1.6 &   0.625 &     0.9841   &  0.6657(17)   \\
0.4 & 1.7 & 0.5882  &      0.6741      &  0.670(7)                 \\
0.6 & 1.8 &  0.5556 &    0.6775   &  0.677(16)   \\
\hline
\end{tabular}\end{center}
\caption{The critical exponents $\nu_C$  for the long-range cubic model at $d=3$ and $N=3$, as computed by a one-loop and three-loop truncation and by a Pad\'e-Borel summation of the three-loop series with $[2/1]$ approximant (with error estimated by the difference with the PB summation of the two-loop series with $[1/1]$ approximant).}
\label{tab:nucubic}
\end{table}

\subsection{The long-range $O(M) \times O(N)$ bifundamental model}
\label{sec:bifundamental}

The last special case of the general multi-scalar model we discuss is the $O(M)\times O(N)$ model. We consider two integers $M$ and $N$, such that $\cN=MN$, and we impose $O(M)\times O(N)$ symmetry. The resulting model, called \emph{bifundamental} in \cite{Rychkov:2018vya}, has two quartic couplings, and it has been extensively studied in its short-range version (e.g.\ \cite{Kawamura:1988,Kawamura:1990,Pelissetto:2001fi,Gracey:2002pm,Delamotte:2003dw,Kompaniets:2020,Henriksson:2020fqi}).

The beta functions have four solutions. The first solution is the trivial one, and the second solution is the Heisenberg fixed-point with  $O(MN)$ symmetry.
The third and fourth solutions are the chiral and anti-chiral fixed-points \cite{Kawamura:1988}. 

There are four regimes of criticality at fixed $M$  depending on the stability of the Heisenberg and chiral fixed points:
\begin{itemize}
\item If $N> N_{c+}$ there are four real fixed points, and the chiral one is stable.
\item If $N_{c-}<N< N_{c+}$, only the Gaussian and the Heisenberg fixed points are real, and they are both unstable.
\item If $N_H <N < N_{c-}$, there are again four real fixed points, and the chiral (or sinusoidal) one is stable.
\item If $N<N_H$, there are still four real fixed points, but the Heisenberg one is stable. 
\end{itemize}

The perturbative expansions  of $N_H$ and $N_{c\pm}$ are given in \cite{Benedetti:2020rrq}.\footnote{Notice that in the published version of \cite{Benedetti:2020rrq} a factor two is missing in front of the $\a_U$ contribution in the formulas. The numerics below take into account the correct factor.}
We can look at their numerical values at $d=3$ and $M=2$. Table \ref{tab:ncp} 
gives values of $N_{c\pm}$ and $N_H$ for different values of $\epsilon$ with either a three-loop truncation or a Pad\'e-Borel summation method. 
The table indicates that for $d=3$ and $M=N=2$ the chiral (or sinusoidal) fixed point might exist and be stable for sufficiently small $\epsilon$. However, at $N=3$ the chiral fixed point is not present, and the Heisenberg one is not stable.  
\begin{table}[H]
\begin{center}
\begin{tabular}{|c|c||c|c|c|}\hline
$\epsilon$ &  & one-loop &        three-loop   &    PB $[1/1]$ \\ \hline
\multirow{3}{*}{0.2} & $N_{c+}$ & 21.8 &   16.14  &  15.6(14) \\
                     & $N_{c-}$ & 2.202 &  2.125  &  2.078(37) \\
                     & $N_{H}$ &   2  &  1.816  &  1.764(16)  \\
\hline
\multirow{3}{*}{0.4} & $N_{c+}$ & 21.8 &   14.43  &  11.2(32) \\
                     & $N_{c-}$ & 2.202 &  2.267  &   2.01(9) \\
                     & $N_{H}$ &   2  &   1.917      &  1.619(40)  \\
\hline
\multirow{3}{*}{0.6} & $N_{c+}$ & 21.8 &  16.69   &  8(5)     \\
                     & $N_{c-}$ & 2.202 & 2.626 &  1.96(15) \\
                     & $N_{H}$ &   2  &  2.302  &  1.52(6) \\
\hline
\end{tabular}\end{center}
\caption{The critical values $N_{c\pm}$ and  $N_{H}$ for the long-range bifundamental model at $d=3$ and $M=2$, as computed by a one-loop and three-loop truncation and by a Pad\'e-Borel summation of the three-loop series with $[1/1]$ approximant (with error estimated by the difference with the PB summation of the two-loop series with $[0/1]$ approximant).}
\label{tab:ncp}
\end{table}

\section*{Acknowledgements}

We are very grateful to Ronnie Rodgers for helping us improve the Mathematica script used to evaluate numerically the infinite sums appearing in our computation. We also thank Benjamin Knorr and Erik Panzer for useful discussions. 

R. G. is supported by the European Research Council (ERC) under the European Union's Horizon 2020 research and innovation program (grant agreement No 818066) and by the Deutsche Forschungs-gemeinschaft (DFG) under Germany's Excellence Strategy EXC--2181/1 - 390900948 (the Heidelberg STRUCTURES Cluster of Excellence).


\addcontentsline{toc}{section}{References}

\providecommand{\href}[2]{#2}\begingroup\raggedright\endgroup



\begin{thebibliography}{10}

\bibitem{Benedetti:2020rrq}
D.~Benedetti, R.~Gurau, S.~Harribey and K.~Suzuki, \emph{{Long-range
  multi-scalar models at three loops}},
  \href{https://doi.org/10.1088/1751-8121/abb6ae}{\emph{J. Phys. A} {\bfseries
  53} (2020) 445008} [\href{https://arxiv.org/abs/2007.04603}{{\ttfamily
  2007.04603}}].

\bibitem{Fisher:1972zz}
M.~E. Fisher, S.-k. Ma and B.~Nickel, \emph{Critical exponents for long-range
  interactions},
  \href{https://doi.org/10.1103/PhysRevLett.29.917}{\emph{Phys.Rev.Lett.}
  {\bfseries 29} (1972) 917}.

\bibitem{Yamazaki:1977pt}
Y.~Yamazaki and M.~Suzuki, \emph{Critical behavior of isotropic systems with
  long range interactions},
  \href{https://doi.org/10.1143/PTP.57.1886}{\emph{Prog. Theor. Phys.}
  {\bfseries 57} (1977) 1886}.

\bibitem{Kotikov:2000yd}
A.~V. Kotikov, \emph{{The Gegenbauer polynomial technique: The Evaluation of
  complicated Feynman integrals}},  in \emph{{15th International Workshop on
  High-Energy Physics and Quantum Field Theory (QFTHEP 2000)}}, pp.~211--217,
  7, 2000, \href{https://arxiv.org/abs/hep-ph/0102177}{{\ttfamily
  hep-ph/0102177}}.

\bibitem{Behan:2023ile}
C.~Behan, E.~Lauria, M.~Nocchi and P.~van Vliet, \emph{{Analytic and numerical
  bootstrap for the long-range Ising model}},
  \href{https://doi.org/10.1007/JHEP03(2024)136}{\emph{JHEP} {\bfseries 03}
  (2024) 136} [\href{https://arxiv.org/abs/2311.02742}{{\ttfamily
  2311.02742}}].

\bibitem{Rong:2024vxo}
J.~Rong, \emph{{Local/Short-range conformal field theories from long-range
  perturbation theory}},  \href{https://arxiv.org/abs/2406.17958}{{\ttfamily
  2406.17958}}.

\bibitem{Li:2024uac}
Z.~Li, \emph{{Conformality loss and short-range crossover in long-range
  conformal field theories}},
  \href{https://arxiv.org/abs/2409.19392}{{\ttfamily 2409.19392}}.

\bibitem{Gorishnii:1984te}
S.~G. Gorishnii and A.~P. Isaev, \emph{{On an Approach to the Calculation of
  Multiloop Massless Feynman Integrals}},
  \href{https://doi.org/10.1007/BF01018263}{\emph{Theor. Math. Phys.}
  {\bfseries 62} (1985) 232}.

\bibitem{Grozin:2012xi}
A.~G. Grozin, \emph{{Massless two-loop self-energy diagram: Historical
  review}}, \href{https://doi.org/10.1142/S0217751X12300189}{\emph{Int. J. Mod.
  Phys. A} {\bfseries 27} (2012) 1230018}
  [\href{https://arxiv.org/abs/1206.2572}{{\ttfamily 1206.2572}}].

\bibitem{Kotikov:2018wxe}
A.~Kotikov and S.~Teber, \emph{{Multi-loop techniques for massless Feynman
  diagram calculations}},
  \href{https://doi.org/10.1134/S1063779619010039}{\emph{Phys. Part. Nucl.}
  {\bfseries 50} (2019) 1} [\href{https://arxiv.org/abs/1805.05109}{{\ttfamily
  1805.05109}}].

\bibitem{Adams:2016xah}
L.~Adams, C.~Bogner, A.~Schweitzer and S.~Weinzierl, \emph{{The kite integral
  to all orders in terms of elliptic polylogarithms}},
  \href{https://doi.org/10.1063/1.4969060}{\emph{J. Math. Phys.} {\bfseries 57}
  (2016) 122302} [\href{https://arxiv.org/abs/1607.01571}{{\ttfamily
  1607.01571}}].

\bibitem{Moch:2001zr}
S.~Moch, P.~Uwer and S.~Weinzierl, \emph{{Nested sums, expansion of
  transcendental functions and multiscale multiloop integrals}},
  \href{https://doi.org/10.1063/1.1471366}{\emph{J. Math. Phys.} {\bfseries 43}
  (2002) 3363} [\href{https://arxiv.org/abs/hep-ph/0110083}{{\ttfamily
  hep-ph/0110083}}].

\bibitem{Mikhailov:2018udp}
S.~V. Mikhailov and N.~I. Volchanskiy, \emph{{Two-loop kite master integral for
  a correlator of two composite vertices}},
  \href{https://doi.org/10.1007/JHEP01(2019)202}{\emph{JHEP} {\bfseries 01}
  (2019) 202} [\href{https://arxiv.org/abs/1812.02164}{{\ttfamily
  1812.02164}}].

\bibitem{Davydychev:1995mq}
A.~I. Davydychev and J.~B. Tausk, \emph{{A Magic connection between massive and
  massless diagrams}},
  \href{https://doi.org/10.1103/PhysRevD.53.7381}{\emph{Phys. Rev. D}
  {\bfseries 53} (1996) 7381}
  [\href{https://arxiv.org/abs/hep-ph/9504431}{{\ttfamily hep-ph/9504431}}].

\bibitem{Glumac:1989}
Z.~Glumac and K.~Uzelac, \emph{{Finite-range scaling study of the 1D long-range
  Ising model}},
  \href{https://doi.org/10.1088/0305-4470/22/20/020}{\emph{J.Phys.A} {\bfseries
  22} (1989) 4439}.

\bibitem{Luijten:1997-thesis}
E.~Luijten, \emph{Interaction Range, Universality and the Upper Critical
  Dimension}. Delft University Press, 1997.

\bibitem{uzelac2001critical}
K.~Uzelac, Z.~Glumac and A.~Ani{\v{c}}i{\'c}, \emph{Critical behavior of the
  long-range ising chain from the largest-cluster probability distribution},
  \href{https://doi.org/10.1103/physreve.63.037101}{\emph{Physical Review E}
  {\bfseries 63} (2001) }.

\bibitem{tomita2009monte}
Y.~Tomita, \emph{Monte carlo study of one-dimensional ising models with
  long-range interactions},
  \href{https://doi.org/10.1143/JPSJ.78.014002}{\emph{Journal of the Physical
  Society of Japan} {\bfseries 78} (2009) 014002}.

\bibitem{Angelini:2014}
M.~C. Angelini, G.~Parisi and F.~Ricci-Tersenghi, \emph{Relations between
  short-range and long-range ising models},
  \href{https://doi.org/10.1103/PhysRevE.89.062120}{\emph{Phys. Rev. E}
  {\bfseries 89} (2014) 062120}
  [\href{https://arxiv.org/abs/1401.6805}{{\ttfamily 1401.6805}}].

\bibitem{Banos:2012}
R.~A. Banos, L.~A. Fernandez, V.~Martin-Mayor and A.~P. Young,
  \emph{Correspondence between long-range and short-range spin glasses},
  \href{https://doi.org/10.1103/physrevb.86.134416}{\emph{Phys. Rev. B}
  {\bfseries B86} (2012) } [\href{https://arxiv.org/abs/1207.7014}{{\ttfamily
  1207.7014}}].

\bibitem{Defenu:2014}
N.~Defenu, A.~Trombettoni and A.~Codello, \emph{{Fixed-point structure and
  effective fractional dimensionality for $O(N)$ models with long-range
  interactions}}, \href{https://doi.org/10.1103/physreve.92.052113}{\emph{Phys.
  Rev. E} {\bfseries 92} (2015) 052113}
  [\href{https://arxiv.org/abs/1409.8322}{{\ttfamily 1409.8322}}].

\bibitem{El-Showk:2014dwa}
S.~El-Showk, M.~F. Paulos, D.~Poland, S.~Rychkov, D.~Simmons-Duffin and
  A.~Vichi, \emph{{Solving the 3d Ising Model with the Conformal Bootstrap II.
  c-Minimization and Precise Critical Exponents}},
  \href{https://doi.org/10.1007/s10955-014-1042-7}{\emph{J. Stat. Phys.}
  {\bfseries 157} (2014) 869}
  [\href{https://arxiv.org/abs/1403.4545}{{\ttfamily 1403.4545}}].

\bibitem{Kleinert:2001ax}
H.~Kleinert and V.~Schulte-Frohlinde, \emph{{Critical properties of
  $\phi^4$-theories}}. World Scientific, River Edge, USA, 2001.

\bibitem{Rychkov:2018vya}
S.~Rychkov and A.~Stergiou, \emph{General properties of multiscalar {RG} flows
  in $d=4-\varepsilon$},
  \href{https://doi.org/10.21468/SciPostPhys.6.1.008}{\emph{SciPost Phys.}
  {\bfseries 6} (2019) 008} [\href{https://arxiv.org/abs/1810.10541}{{\ttfamily
  1810.10541}}].

\bibitem{Kawamura:1988}
H.~Kawamura, \emph{Renormalization-group analysis of chiral transitions},
  \href{https://doi.org/10.1103/PhysRevB.38.4916}{\emph{Phys. Rev. B}
  {\bfseries 38} (1988) 4916}.

\bibitem{Kawamura:1990}
H.~Kawamura, \emph{Generalized chiral universality},
  \href{https://doi.org/10.1143/JPSJ.59.2305}{\emph{Journal of the Physical
  Society of Japan} {\bfseries 59} (1990) 2305}.

\bibitem{Pelissetto:2001fi}
A.~Pelissetto, P.~Rossi and E.~Vicari, \emph{{Large $n$ critical behavior of
  $O(n) \times O(m)$ spin models}},
  \href{https://doi.org/10.1016/S0550-3213(01)00223-1}{\emph{Nucl. Phys. B}
  {\bfseries 607} (2001) 605}
  [\href{https://arxiv.org/abs/hep-th/0104024}{{\ttfamily hep-th/0104024}}].

\bibitem{Gracey:2002pm}
J.~Gracey, \emph{{Chiral exponents in $O(N) \times O(m)$ spin models at $O(1 /
  N^2)$}}, \href{https://doi.org/10.1103/PhysRevB.66.134402}{\emph{Phys. Rev.
  B} {\bfseries 66} (2002) 134402}
  [\href{https://arxiv.org/abs/cond-mat/0208309}{{\ttfamily
  cond-mat/0208309}}].

\bibitem{Delamotte:2003dw}
B.~Delamotte, D.~Mouhanna and M.~Tissier, \emph{{Nonperturbative
  renormalization group approach to frustrated magnets}},
  \href{https://doi.org/10.1103/PhysRevB.69.134413}{\emph{Phys. Rev. B}
  {\bfseries 69} (2004) 134413}
  [\href{https://arxiv.org/abs/cond-mat/0309101}{{\ttfamily
  cond-mat/0309101}}].

\bibitem{Kompaniets:2020}
M.~Kompaniets, A.~Kudlis and A.~Sokolov, \emph{{Six-loop $\epsilon$ expansion
  study of three-dimensional $O(n)\times O(m)$ spin models}},
  \href{https://doi.org/https://doi.org/10.1016/j.nuclphysb.2019.114874}{\emph{Nuclear
  Physics B} {\bfseries 950} (2020) 114874}.

\bibitem{Henriksson:2020fqi}
J.~Henriksson, S.~R. Kousvos and A.~Stergiou, \emph{{Analytic and Numerical
  Bootstrap of CFTs with $O(m)\times O(n)$ Global Symmetry in 3D}},
  \href{https://arxiv.org/abs/2004.14388}{{\ttfamily 2004.14388}}.

\end{thebibliography}
\end{document}